\documentclass[]{aastex}
\usepackage{txfonts,amssymb,graphicx,color}
\begin{document}

\title{REVIEW: A COHERENT AND COMPREHENSIVE MODEL OF THE EVOLUTION OF THE
  OUTER SOLAR SYSTEM}

\author{Alessandro Morbidelli}
\affil{\small\em Departement Cassiop\'{e}e: Universit\'e de
Nice - Sophia Antipolis, Observatoire de la C\^{o}te d'Azur, CNRS.
06304 Nice Cedex 4, France}

\received{by CRAS, on Sept 20, 2010}
\accepted{by CRAS, on Oct 27, 2010}

\begin{abstract}

Since the discovery of the first extra-solar planets, we are
confronted with the puzzling diversity of planetary systems. Processes
like planet radial migration in gas-disks and planetary orbital
instabilities, often invoked to explain the exotic orbits of the
extra-solar planets, at first sight do not seem to have played a role
in our system. In reality, though, there are several aspects in the
structure of our Solar System that cannot be explained in the classic
scenario of in-situ formation and smooth evolution of the giant
planets. This paper describes a new view of the evolution of the outer
Solar System that emerges from the
so-called 'Nice model' and its recent extensions.  The story provided
by this model describes a very ``dynamical'' Solar System, with giant
planets affected by both radial migrations and a temporary orbital
instability. Thus, the diversity between our system and those found so
far around other stars does not seem to be due to different processes
that operated here and elsewhere, but rather stems from the strong
sensitivity of chaotic evolutions to small differences in the initial
and environmental conditions.

\vskip 20pt

In press in ``C.R. Physique de l'Acad\'emie des Sciences''.

\end{abstract}

\section{Introduction}

When looking at the structure of the outer Solar System, i.e. the four
giant planets and the populations of small bodies from the orbit of
Jupiter outwards, one sees several puzzling aspects that do not fit the
simple scenario of in-situ formation of planets from a circum-solar
disk of gas and solids, developed over the last centuries from the
ideas of Laplace (Laplace, 1796).  Moreover, our Solar System looks
quite different from the planetary systems discovered so far around
other stars.

For instance: (i) many extra-solar giant planets have small orbital
radii, comparable (or smaller) than those of the terrestrial planets
of our Solar System (Mercury to Mars); instead, our giant planets
(Jupiter to Neptune) orbit the Sun at a distance of 5-30 times that of
the Earth. (ii) Giant planets are expected to form on circular and
co-planar orbits; however, the orbital eccentricities and inclinations
of our giant planets, although small, are definitely much larger than
expected from formation models; the orbits of the majority of the
extra-solar giant planets are even more at odds with the theoretical
expectations, because they are much more eccentric than the orbits of
the planets of our system. (iii) Many extra-solar systems have planets
in mutual mean motion resonances, where the ratio of the orbital
periods is equal to a ratio of small integer numbers (often 1/2); but
the orbits of the planets of our system do not have this
property. (iv) One would expect to find, beyond the orbit of the last
planet, a disk of small icy objects, called planetesimals, that
preserves its original, virgin structure: quasi-circular and coplanar
orbits and a cumulatively large total mass; instead the Kuiper belt
(the population of icy bodies tha have been found beyond the orbit of
Neptune) is in total less massive than our Moon, it has an abrupt
outer edge at the location of the 1/2 resonance with Neptune and the
eccentricities and inclinations of its objects can be as large as
allowed by stability constraints. (v) One would expect that the Solar
System evolved gradually, from a primordial chaos characterized by
mutual collisions and ejections of bodies, to the current state of
essentially regular orbital motion; however, the terrestrial planets,
the asteroids and, possibly, the satellites of the giant planets,
carry the scars of a ``Late Heavy Bombardment'' (LHB), suddenly
triggered 600 million years after planet formation, or approximately
3.9 Gy (Giga-year) ago. This argues for a sudden change in the
structure of the Solar System, so that a stable reservoir of small
bodies became unstable and its objects started to intersect those of
the planets and collide with the latter.

The 'Nice model' -so named because it was developed at the
Observatoire de la C\^ote d'Azur in Nice- has the ambition to explain
all these and other intriguing features in the framework of a unitary
scenario. Several other models have been developed over the years to
explain one or another of the puzzling properties of our Solar System,
but none has the comprehensive character of the Nice model. 

In this paper, I will review the basic ideas behind this model.  In
section 2, I will present the original version of the model, as
proposed in 2005. The model has vastly evolved since then, in order to
overcome its limitations and extend the time-span of the events that
it can describe. I will discuss these evolutions in Section~3. Section
4 will summarize the current view of Solar System evolution that
emerges from this model.

\section{The original model}

The original Nice model was developed to explain the origin
of the small, but non-negligible eccentricities and inclinations of the
giant planets and the origin of the Late Heavy Bombardment of the
inner Solar System. 

Like most models, the Nice model was based on pre-existing ideas.
First, it was known since Fernandez and Ip (1984) that, after the
disappearance of the gas, while scattering away the primordial
planetesimals from their neighboring regions, the giant planets had to
migrate in semi-major axis as a consequence of angular momentum
conservation.  Given the configuration of the giant planets in our
Solar System, this migration should have had a general trend. Uranus
and Neptune have difficulty ejecting planetesimals onto hyperbolic
orbits.  Apart from the few percent of planetesimals that they can
permanently store in the Oort cloud (the shell-like reservoir of
long-period comets, situated at about $10^4$ Astronomical Units (AU)
from the Sun; Dones et al., 2004), or emplace onto long-lived orbits
in the trans-Neptunian region (Duncan and Levison, 1997), the large
majority of the planetesimals that are under the influence of Uranus
and Neptune are eventually scattered inwards, towards Saturn and
Jupiter.  Thus, Uranus and Neptune, by reaction, have to move
outwards. Jupiter, on the other hand, eventually ejects from the Solar
System almost all of the planetesimals that it encounters: thus it has
to move inwards. The fate of Saturn is more difficult to predict, a
priori. However, modern numerical simulations show that this planet
also moves outwards, although only by a few tenths of an AU for
reasonable disk's masses (e.g. $\sim 50$ Earth masses; see Hahn and
Malhotra, 1999; Gomes et al., 04).

Second, it was known that planets embedded in a planetesimal disk
suffer ``dynamical friction'' which damps their orbital eccentricities
and inclinations (Wetherill and Stewart, 1993). Thus, the planetesimal
scattering process that leads to planet migration by itself cannot
enhance the eccentricities and inclinations of the planets relative to
their (almost null) initial values (Morbidelli et al., 2009). However, it was
also known that, if the planets cross mutual mean motion resonances
during their divergent migration, their eccentricities are enhanced
almost impulsively (Chiang, 2003).  The eccentricity increase depends
on the planetary masses and on the resonance involved. If the
eccentricities become too large, then planets can become unstable.

Third, it was shown by Thommes et al. (1999) that the instability of
the giant planets system would not necessarily lead to the disruption
of the outer Solar System. In several cases, Uranus and Neptune are
scattered outwards by Jupiter and Saturn; then the interaction with the
disk of planetesimals can damp by dynamical friction the eccentricities of
Uranus and Neptune, preventing them to have further close encounters with
Jupiter or Saturn and between themselves; consequently, the 4-planet
system can achieve a new stable configuration.   

Last, Levison et al. (2001) showed that the dispersal of a planetesimal
disk of $\sim 50$ Earth masses by the migrating giant planets 
would induce a bombardment of the
terrestrial planets of magnitude comparable to that of the LHB; thus
the problem of the origin of the LHB is re-conduced to the problem of
finding a plausible mechanism for triggering giant planet migration at
a correspondingly late time. \footnote{It was proposed in
Levison et al. (2001) that this mechanism was the late formation of Uranus
and Neptune, but a formation as late as 600~My is inconsistent with
the physical structure of these planets (which contain hydrogen and
helium in roughly solar proportion) and with their dynamics during
accretion (Levison et al., 2007).}

\begin{figure}
\includegraphics[height=13cm]{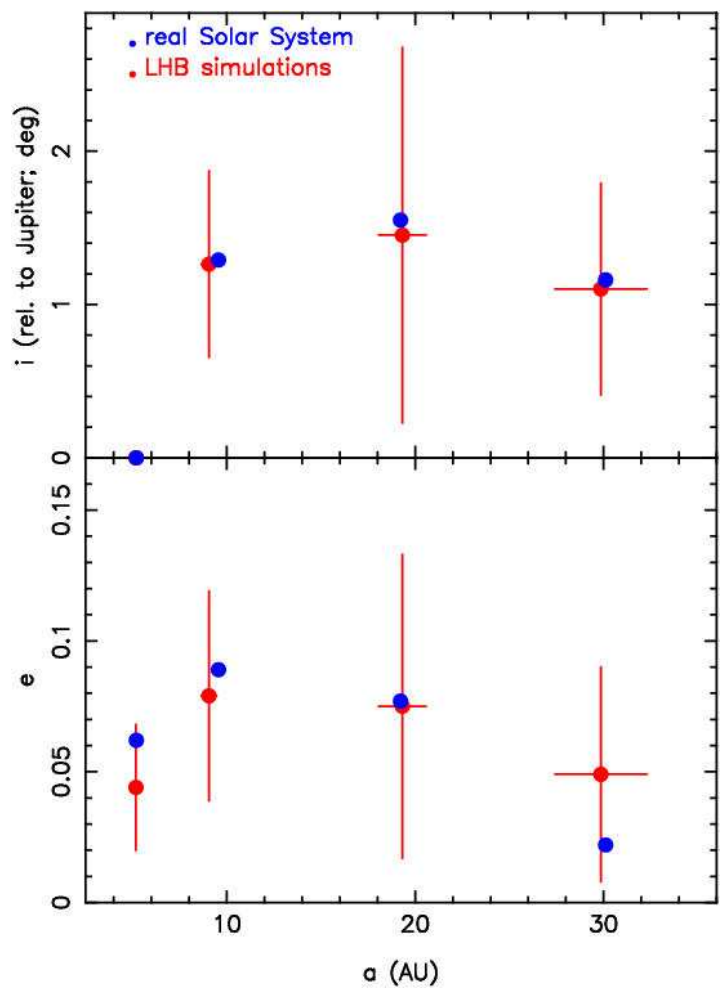}
\caption{Comparison of the synthetic final planetary systems obtained
in Tsiganis et al. (2005) with the real outer Solar System. Top:
Proper eccentricity vs.  semi-major axis. Bottom: Proper inclination
vs.  semi-major axis.  Here, proper eccentricities and inclinations
are defined as the maximum values acquired over a 2~My time-span and
were computed from numerical integrations. The inclinations are
measured relative to Jupiter’s orbital plane.  The values for the real
planets are presented as filled blue dots.  The red dots mark the
mean of the proper values for 15 simulations. The error bars represent
one standard deviation of the measurements.  }
\label{Tsiganis}
\end{figure}

Building on all these results, the Nice model postulated that, at the
time of the dissipation of the gas-disk, the four fully-grown giant
planets were in a compact configuration, with quasi-circular, coplanar
orbits (as predicted by planet formation models) and with orbital
radii ranging from 5.5 to 17 AU; Saturn and Jupiter were close enough
to each other to have a ratio of orbital periods smaller than 2
 (Tsiganis et al., 2005; the current ratio of their orbital periods is almost
2.5).  During their planetesimal-driven divergent migration, Saturn
and Jupiter increased their orbital period ratio. Thus, with the
adopted initial conditions, Saturn and Jupiter eventually crossed
their mutual 1/2 mean-motion resonance (which occurs when the period
ratio is exactly 2). This resonance enhances the eccentricities of
Jupiter and Saturn, enough to make the whole 4-planet system
unstable. The dynamics then evolves through mutual scattering among
the planets and dynamical friction exerted by the disk, as described
above.  Eventually a new stable configuration is achieved once all
disk particles are dispersed and removed.  The simulations in
Tsiganis et al. (2005) show that, if the planetesimal disk contained about 35
Earth masses and was truncated at $\sim 35$ AU, this dynamical
evolution leads to a final orbital configuration of the planetary system that
reproduces the current configuration remarkably well, in terms of
semi-major axes, eccentricities and inclinations (see
Fig.~\ref{Tsiganis}).

With this result in hands, Gomes et al. (2005) could put all the
elements together in a coherent scenario for the LHB origin.  They
reasoned that, at the end of the gas-disk phase, the planetesimal disk
should have contained only those bodies that had dynamical lifetimes
longer than the lifetime of the solar nebula (a few million years),
because the planetesimals initially on orbits with shorter dynamical
lifetimes should have been eliminated earlier, during the nebula era.
Assuming the initial planetary system of Tsiganis et al. (2005), this
constraints the planetesimal disk to start about 1 AU beyond the
position of the last planet. With this kind of disk, the 1/2 resonance
crossing event that destabilizes the planetary system occurs in
the simulations of Gomes et al. (2005) at a time ranging from 192 My to 875
My. Modifying the initial planetary orbits also leads to changes in the
resonance crossing time, pushing it up to 1.1 Gy after the beginning
of the simulation. This range of instability times brackets well the
date of the LHB, as estimated from lunar data.

\begin{figure}
\includegraphics[height=7cm]{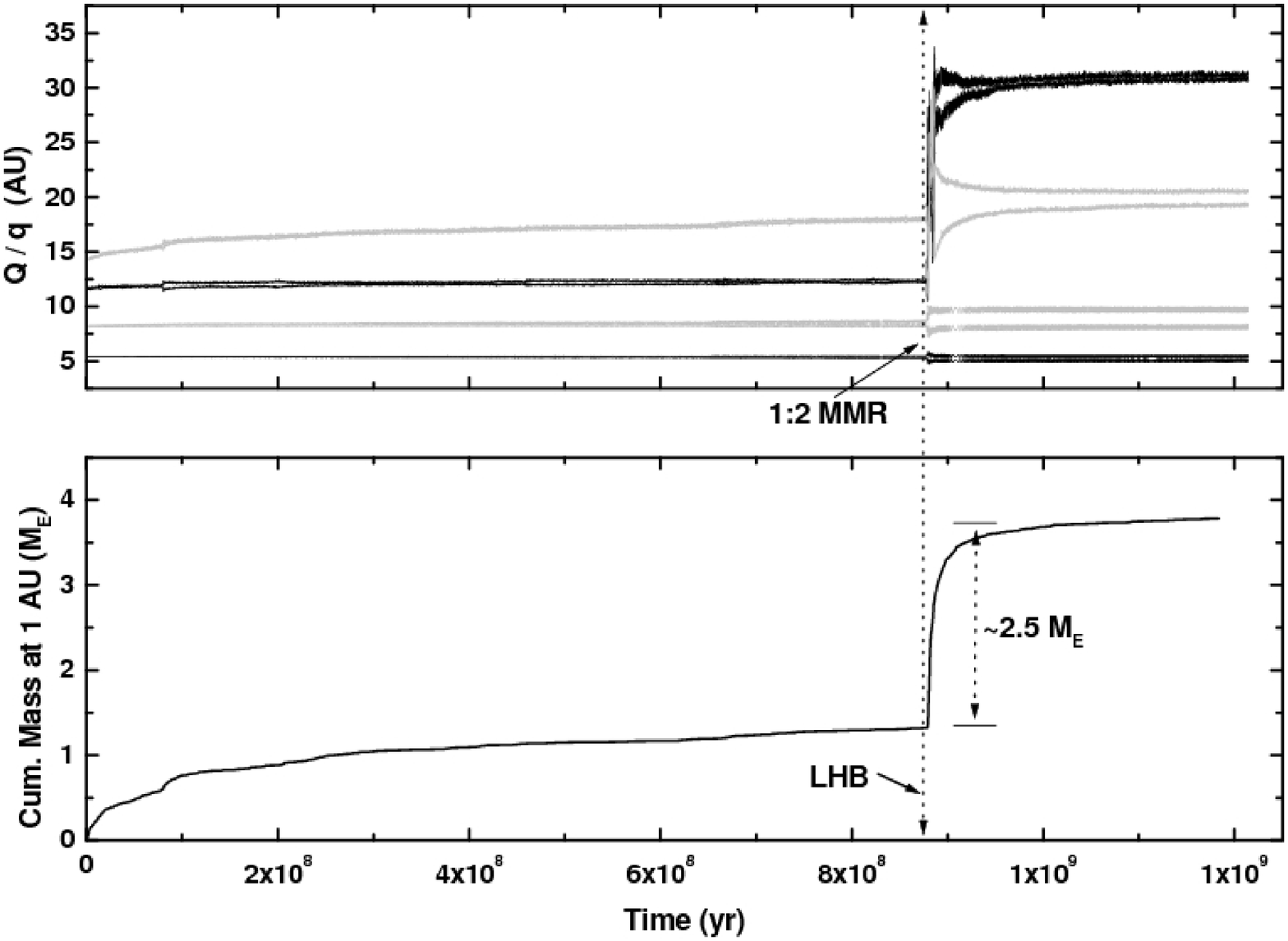}
\caption{Planetary migration and the corresponding mass flux towards the
inner Solar System, from a representative simulation of Gomes et al. (2005).
Top: the evolution of the 4 giant planets.  Each planet is represented
by a pair of curves - the aphelion and perihelion distances.  In this
simulation Jupiter and Saturn cross their 1/2 mean-motion resonance at
880~My.  Bottom: the cumulative mass of comets (solid curve) and
asteroids (dashed curve) accreted by the Moon.  The comet curve is
offset so that the value is zero at the time of 1/2 resonance
crossing.}
\label{Gomes}
\end{figure}

The top panel of Fig.~\ref{Gomes} shows the giant planets' evolution
in a representative simulation of Gomes et al. (2005). Initially, the
giant planets migrated slowly due to the leakage of particles from the
disk. This phase lasted 875 My, at which point Jupiter and Saturn
crossed their 1/2 resonance.  At the resonance crossing event, as in
Tsiganis et al. (2005), the orbits of the ice giants became unstable and they
were scattered into the disk by Saturn.  They disrupted the disk and
scattered objects all over the Solar System, including the inner
regions. Eventually they stabilized on orbits very similar to the
current ones, at $\sim$20 and $\sim$30 AU respectively. The solid
curve in the bottom panel shows the amount of material from the
primordial trans-Neptunian disk that struck the
Moon, as a function of time.  The amount of material hitting the Moon
after the resonance crossing event is consistent with the mass ($6\times
10^{21}$g) estimated from the number and size distribution of lunar
basins that formed around the LHB epoch (Levison et al., 2001).

However, the planetesimals from the distant disk -which can be
identified as `comets'- were not the only ones to hit the terrestrial
planets. The radial migration of Jupiter and Saturn forced secular
resonances (resonances between the precession periods of the asteroids
and of the giant planets) to sweep across the asteroid belt, exciting
the eccentricities and the inclinations of asteroids.  The fraction of
the main belt population that acquired planet-crossing eccentricities
depends quite crucially on the orbital distribution that the belt had
before the LHB, which is not well known.  According to the simulations
in O'brien et al. (2007), at the end of the terrestrial planet formation
process, which pre-dates the LHB, the asteroid belt should have had a
dynamical excitation comparable, or slightly larger than the current
one. In these conditions of orbital excitation, the secular resonance
sweeping at the time of the LHB would have left $\sim$5-10\% of the
objects in the asteroid belt (Gomes et al., 2005). Thus, at the LHB time,
the asteroid belt would have been 10-20 times more massive than
now. In this case, the total mass of the asteroids hitting the Moon
would have been comparable to that of the comets (see
Fig.~\ref{Gomes}).

\subsection{Other successes of the Nice model}

To validate or reject a model, it is important to look at the largest
possible number of constraints. Three populations immediately come to
mind when considering the Nice model: the Trojans and the satellites
of the giant planets and the Kuiper belt. Are their existence and structure
consistent with the Nice scenario?

\vskip 10pt
\noindent{TROJANS}

Jupiter and Neptune have a conspicuous populations of Trojan
objects. These bodies, usually referred to as `asteroids', follow
essentially the same orbit as the planet, but lead or trail that
planet by an angular distance of $\sim\!60$ degrees, librating around
the Lagrange triangular equilibrium points. The latter are the two
positions where a small object, affected only by the gravity of the Sun and
of one planet, can be stationary and stable relative to two larger
objects; together with the positions of the Sun and the planet, they
form two equilateral triangles, rotating in space (Lagrange, 1787).

To date, the number of known Jupiter Trojans is 4526. Probably all
those larger than about 20km in diameter are now known; they are about
1,000 objects.  Instead, only seven Trojan of Neptune are now known,
but detection statistics imply that the Neptune Trojan population
could be comparable in number to that of Jupiter, and possibly even
ten times larger (Chiang and Lithwick, 2005).

The simulations in Tsiganis et al. (2005) and Gomes et al. (2005) led
to the capture of several particles on long-lived Neptunian Trojan
orbits (2 per run, on average, with a lifetime larger than
80~My). Their eccentricities, during their evolution as Trojans,
reached values smaller than $0.1$.  These particles were eventually
removed from the Trojan region, but this is probably an artifact of
the graininess of Neptune's migration in the simulation, due to the
quite large individual mass of the planetesimals.

Jovian Trojans are a more subtle issue that was addressed in detail in
Morbidelli et al. (2005). There is a serious argument in the
literature against the idea that Jupiter and Saturn crossed their 1/2
mean-motion resonance: if the crossing had happened, any pre-existing
Jovian Trojans would have become violently unstable, and Jupiter's
co-orbital region would have emptied (Gomes, 1998; Michtchenko et al.,
2001). However, the dynamical evolution of a gravitating system of
objects is time reversible. Thus, if the original objects can escape
the Trojan region when it becomes unstable, other bodies can enter the
same region and be temporarily trapped.  Consequently, a transient
Trojan population can be created if there is an external source of
objects.  In the framework of the Nice model, the source consists of
the very bodies that are forcing the planets to migrate, which must be
a large population given how far the planets must migrate.  When
Jupiter and Saturn move far enough from the 1/2 resonance that the
co-orbital region becomes stable, the population that happens to be
there at that time remains trapped. It then becomes the population of
permanent Jovian Trojans still observable today.

This possibility has been tested with numerical simulations in
Morbidelli et al. (2005). It was shown that the population of captured Trojans
is consistent, in terms of total mass and orbital distribution, with
the real population.  In particular, the Nice model is the only model
proposed so far which explains the inclination distribution of the
Jovian Trojans. The origin of this distribution was considered to be
the hardest problem in the framework of the classical scenario,
according to which the Trojans formed locally and were captured at the
time of Jupiter's growth (Marzari et al., 2002).

\vskip 10pt
\noindent{IRREGULAR SATELLITES}

The known irregular satellites of the giant planets are dormant
comet-like objects that reside on stable prograde and retrograde
orbits at large distances from the central object, where planetary
perturbations are only slightly larger than solar ones.

One particularity of the irregular satellite systems is that, once the
orbital radii are scaled relative to the radius of the sphere of
gravitational influence of the respective planets, they are all very
similar to each other (Jewitt and Sheppard, 2002). This invalidates
the most popular models proposed for their origin, i.e. (i) capture
due to the sudden growth of the giant planets (Heppenheimer and Porco,
1977) and (ii) capture due to gas drag in the primordial extended
atmosphere of the giant planets (Cuk and Burns, 2004; Kortenkamp, 2005).  In
fact, Jupiter and Saturn are very different from Uranus and Neptune:
presumably the former grew much faster and had much more gas in their
extended atmospheres than the latter, which are essentially gas-poor,
ice-giant planets. Thus, one would expect that both processes (i) and
(ii) would have led to substantially different irregular satellite
systems for these two pairs of planets.

The capture of irregular satellites in the framework of the Nice model
has been investigated in details in Nesvorny et al. (2007). A
characteristic feature of the Nice model is that, at the instability
time, the giant planets suffered mutual close encounters. It was
proposed in Nesvorny et al. (2007) that planetesimals wandering in the vicinity
of the sites of such encounters could become trapped onto permanent
orbits around the planets via gravitational three-body reactions.
Numerical simulations showed that this process is effective, and
leads to orbital distributions of satellites very similar to those
observed around each planet. Moreover, assuming that the mass of the
planetesimal disk was as in the Nice model and that the planetesimals
had a size distribution similar to today's Kuiper belt, the capture
efficiencies predict quite correctly the sizes of the largest
irregular satellites around each planet. The current size distribution
of irregular satellites, which is much shallower than that of the
Kuiper belt, is then explained by their post-capture intense
collisional evolution (Bottke et al., 2010).

From the results in Nesvorny et al. (2007), the capture of the irregular
satellites of Saturn, Uranus and Neptune is a generic process because
these planets experience planet-planet encounters in all the
successful simulations of the Nice model. Instead, the capture of the
satellites of Jupiter is not generic, because in most realization of
the Nice model Jupiter does not encounter another planet. Only some of
the successful simulations of the Nice model have Jupiter-Uranus or
Jupiter-Neptune encounters. The fact that Jupiter has an irregular
satellite system like that of the other planets argues that such
encounters did happen in reality.

This conclusion is
supported by the investigation of the orbital evolution of the
terrestrial planets (Brasser et al., 2009) and of the asteroid belt
(Morbidelli et al., 2010). These studies show that, in absence of
encounters between Jupiter and another planet, the orbital separation
between Jupiter and Saturn would have increased slowly and,
consequently, the orbits of the terrestrial planets would have
acquired too large eccentricities and the final orbital distribution
in the asteroid belt would have become inconsistent with that
observed. Instead, if Jupiter had had an encounter with an ice giant,
the orbital separation between Jupiter and Saturn would have increased
impulsively; this would have allowed the terrestrial planets to stay
on moderate eccentricity orbits and the asteroid belt to avoid the
formation of spurious empty regions within its boundaries.

\vskip 10pt
\noindent{THE KUIPER BELT}

In the Nice model, the proto-planetary disk is assumed do have an
outer edge at about 35 AU, otherwise Neptune ends its migration too
far from its current orbital position. Proto-planetary disks often
have sharp outer edges, as inferred from the radial distribution of
dust in debris disks (the disk of AU Mic, for instance, is inferred to
have an outer edge at 30 AU; Augereau andBeust, 2006). These edges
might have been formed by several mechanisms, such as tidal truncation
during early close stellar flybys (Kenyon an Bromley, 2004),
photo-evaporation of the outer part of the proto-planetary disk (Adams
et al., 2004), ineffective planetesimal accretion where the solid/gas
ratio is too low (Youdin and Goodman, 2005). Thus, it is reasonable to
assume that the planetesimal disk of the Solar System had an outer
edge, but the assumption that this edge was at 35 AU seems to be in
conflict with the existence of a Kuiper belt between 35 and 50 AU. If
the Nice model is correct, then there must be a mechanism to fill with
objects an initially empty Kuiper belt.

Such mechanism was identified by Levison et al. (2008) with the temporary
large eccentricity phase of Neptune at the time of the planetary
instability.  The point is that, when Neptune's orbit is eccentric,
the full $(a,e)$ region up to the location of the 1/2 resonance with
the planet is chaotic. Thus, we can envision the following scenario.
Assume, in agreement with several of the simulations of the Nice
model, that the large eccentricity phase of Neptune is achieved when
the planet has a semi-major axis of $\sim 28$~AU, after its last
encounter with Uranus. In this case, a large portion of the current
Kuiper belt is already interior to the location of the 1/2 resonance
with Neptune. Thus, it is unstable, and can be invaded by objects
coming from within the outer boundary of the disk (i.e. within $\sim
35$~AU).  When the eccentricity of Neptune damps out, the mechanism
for the onset of chaos in the Kuiper belt region disappears. The
Kuiper belt becomes stable, and the objects that happen to be there at
that time remain trapped for the eternity.

The simulations of Levison et al. (2008) successfully implanted a small
fraction (approximately 1/1000) of the disk's planetesimals into the
current Kuiper belt. This explains the low mass of the observed Kuiper
belt population. The major success of the simulations is to reproduce
the current sharp outer edge of the Kuiper belt, located at the
position of the 1/2 resonance with Neptune. This is the first, and so
far only model, capable of explaining this characteristic of the
belt. The observed orbital distribution in the Kuiper belt is also
fairly well reproduced in the simulations, although the match is not
perfect. For instance, there is a deficit in the synthetic population
above 20 degrees of inclination.

\section{The new Nice model}

Despite of its successes, the original Nice model has some important
weaknesses. The most important one is that the initial orbits of the
giant planets are totally arbitrary. The assumption of small
eccentricities and inclinations is reasonable, as this is expected
from planet formation models, but the original orbital semi major axes
are totally made up. In particular, Saturn and Jupiter are set
initially on orbits close to their mutual 1/2 resonance. The initial
distance from this resonance is more or less tuned so to have an
instability around the LHB time. Had this distance been larger, the
planets would have not reached the resonance and would not have become
unstable; had this distance been smaller, the resonance crossing would
have occurred too early.

Clearly, there is the need to justify better the initial orbits of the
planets. The initial conditions of the Nice model should correspond to
the orbital structure that the Solar System had when it emerged from
the gas-disk phase. Thus, the best way to set a valid initial orbital
configuration of the planets is to study the dynamical evolution that
said planets should have had when they were still embedded in the gas.  

\subsection{The dynamics of the giant planets in a gas-disk}

The gravitational interaction of planets with a disk of gas leads to
the orbital migration of the former, on a timescale and a radial range
respectively much shorter and much wider than those characterizing the
migration induced by the interaction with the planetesimals in a
gas-less disk. The gas-driven migration is named ``Type-I'' for medium-mass
planets like Uranus and Neptune that do not open a gap in the gas-disk
around their orbits; it is named ``Type-II'' for giant planets like Jupiter and
Saturn that do open at least partial gaps. Both migrations generically
force the planetary orbits to shrink. The discovery of a large number
of extra-solar giant planets on orbits with small radii (less than 1
AU; even less than 0.1 AU in the case of  the so-called ``Hot
Jupiters'') is an empirical demonstration that radial migration occurs
in real nature. 

As stated in the introduction of this paper, whoever studies planet
migration in gas-disks is confronted with crucial questions: why is
Jupiter at 5 AU? Why did Jupiter not migrate closer to the Sun, unlike
most of the known extra-solar planets?

\begin{figure}[t!]
\centerline{\includegraphics[height=6.cm]{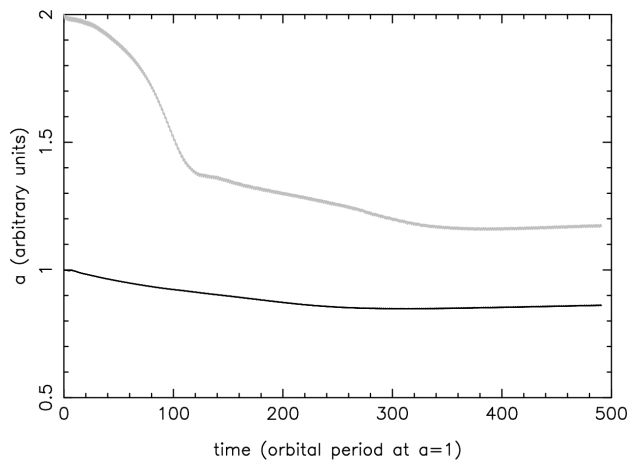}}
\vspace*{-.3cm} 
\caption{\small An illustration of the dynamical evolution of Jupiter
  and Saturn in the gas-disk, as in Masset and Snellgrove (2001). The
  black and grey curves show the evolutions of the semi major axes of
  Jupiter and Saturn, respectively. Capture in the 2/3 mean motion
  resonance occurs when the migration of Saturn is reversed.}
\label{MS01} 
\end{figure}

The answer  relies
on the co-existence of Jupiter and Saturn, with their specific 3:1
mass ratio. In fact, as first showed in Masset and Snellgrove (2001) with 
hydro-dynamical simulations, Saturn migrates inwards faster than Jupiter and
consequently it approaches the major planet until it is trapped in
its 2/3 mean motion resonance (where the orbital period of
Jupiter is 2/3 that of Saturn; see Fig~\ref{MS01}). More recently,
it has been shown (Pierens and Nelson, 2008) that the capture of a Saturn-like
planet into the 2/3 resonance with a Jupiter-like planet is a very
robust outcome of simulations, independent of initial conditions and
of the mass-growth history of the outer planet.

Once in 2/3 resonance configuration, the planets cease migrating
inwards. It was shown in Morbidelli and Crida (2007) that the subsequent
orbital evolution depends on the properties of the disk, particularly
the scale height. In general, both planets migrate outwards together,
on a short timescale. However, if the disk is very thick, the
migration rate is very slow, as in Fig~\ref{MS01}. For some
appropriate disk thickness there is essentially no migration. 
\footnote{Notice that, for two giant planets to avoid inward migration
by this mechanism, it is essential that the mass of the outer planet
is a fraction of the mass of the inner planet, as in the
Jupiter-Saturn case (Masset and Snellgrove, 2001; Morbidelli and
Crida, 2007). Planets of
comparable masses or with a reversed mass ratio do migrate towards the
central star also after resonance trapping.}

The presence of asteroids inside of Jupiter's orbit suggests at first
sight that Jupiter never came closer to the Sun than its present
position. Thus the parameters of the circum-solar disk should have
been close to those resulting in a non-migrating evolution of the
Jupiter-Saturn pair after their trapping in resonance. However, it has
been proposed (Walsh et al., 2010) that Jupiter migrated down to 1.5 AU before
Saturn formed and was captured in resonance; when this occurred, the
two planets reversed migration and Jupiter reached 5.4 AU when the
gas-disk disappeared. It has been argued that this kind of evolution
would explain the properties of the terrestrial planets -in particular
the large Earth/Mars mass ratio- and of the asteroid belt -in
particular the dichotomy of physical properties of inner belt
vs. outer belt asteroids- better than any other existing model. In
either case, the 2/3 resonance configuration of Saturn and Jupiter
explains why Jupiter did not come to, or did not stay at, a small
distance from the Sun.

The presence of the two major planets in a configuration characterized by
no inward migration must have strongly influenced the dynamics of
Uranus and Neptune and may even have played a role in their
accretion. In fact, any sizeable proto-planet formed in the outer
disk should have migrated inwards by Type I migration, until being
trapped in some resonance with Saturn, or at the outer edge of its gap
(Pierens and Nelson, 2008). The accumulation of embryos at specific
sites outside the orbit of Saturn may have boosted the accretion of
the cores of Uranus and Neptune. This phase, however, has never been
modeled in details.

A search for possible orbital configurations of Uranus and Neptune
relative to Jupiter and Saturn was done in Morbidelli et al. (2007),
with a step-wise approach. First Jupiter and Saturn were set in a 2/3
resonant, non-migrating orbital configuration. Then several
hydro-dynamical simulations were done, placing Uranus at various
orbital separations from Saturn and assuming a disk density close to
the so-called ``Minimum Mass Solar Nebula'' (Weidenschilling,
1977). It was observed that Uranus migrated too fast to be trapped in
the 1/2 resonance with Saturn. Conversely it could be trapped,
depending on the initial conditions, in the 2/3 or 3/4
resonances. Configurations with Uranus closer to Saturn than the 3/4
resonance turned out to be unstable, with Uranus chased outwards by a
distant encounter with Saturn, and eventually trapped in one of the
two resonances listed above. Finally, for each of the two final stable
configurations achieved by Uranus, a second set of hydro-dynamical
simulations was done placing Neptune at various initial orbital
separations from Uranus. It was observed that Neptune migrated too
fast to be trapped in either the 1/2 or 2/3 resonances with
Uranus. Instead, it could be trapped, depending on initial conditions,
into the 3/4, 4/5 or 5/6 resonances. Thus, in total 6 orbital
configurations could be found, in which all planets are in resonance
with each other. Other possible multi-resonant configurations of the
giant planets have been found by Batygin and Brown (2010) using N-body
integrations with forces that mimic Type-I migration of Uranus and
Neptune.

\subsection{The dynamics of the giant planets after the gas-disk removal}

Many of the multi-resonant configurations described above are stable
on Gy timescale once the gas-disk is removed.  However, if there is a
remnant planetesimal disk, the planet-planetesimals interactions
perturb the orbits of the planets, and eventually may extract the
planets from their mutual resonances.  Resonances have a strong
stabilizing effect for close orbits (a clear example is that of Pluto
which, despite it crosses the orbit of Neptune, is stable because it
is in its 2/3 resonance). Once the planets are extracted from their
mutual resonances, this stabilizing effect ends. The planets rapidly
become unstable, because they are too close to each other. A phase of
mutual scattering starts, similar to that described in the original
Nice-model paper (Tsiganis et al., 2005; Gomes et al., 2005).  The
simulations in Morbidelli et al. (2007) and Batygin and Brown (2010)
show that the final orbits that the planets achieve once the
planetesimal disk is dispersed are often similar to the real
ones. This shows that the multi-resonant configuration, which the
giant planets should have been driven into during the gas-disk phase,
can be consistent with the current orbital architecture of the
planets, provided that the latter passed through a global instability
phase.

Could this instability occur late, as in the original Nice model
(Gomes et al., 2005), so to explain the origin of the LHB?  A delayed
instability can not be simply obtained by assuming that the
planetesimal disk starts approximately $1$~AU beyond the orbit of the
furthermost planet, as in Gomes et al. (2005). In fact, the planets are
now in resonances with each other, and the combination of resonance
locking among the planets with the planet-planetesimal scattering
process makes the instability time much more sensitive to the exact
location of the disk's inner edge than in Gomes et al. (2005). Such an
extreme sensitivity to the disk's parameters is, of course,
problematic.

This problem, however, appears only in simulations which,
like all those of the papers quoted above, assume that the
planetesimals do not interact dynamically with each other. Instead,
if self-interactions are taken into account, for instance assuming
that there are a few 100s Pluto-mass objects in the disk perturbing
each other and the other particles, then there is a net exchange of
angular momentum between the planets and the disk, even if there are
no close encounters between planets and planetesimals. In particular,
the planets loose energy and momentum, i.e. they try to migrate
towards the Sun (Levison et al., 2011).  The orbits of the planets tend
to {\it approach} each other. This is different from the case where
planets scatter planetesimals, in which the planetary orbits tend to
{\it separate} from each other. Remember, though, that the planets are
in resonances; so the ratios between their semi major axes cannot
change. In response, the planetary eccentricities slowly {\it
increase}. This eventually drives some planets to pass through
secondary or secular resonances, which destabilize the original
multi-resonant configuration. Due to this process, the instability
time is late in general: in the simulations of Levison et al. (2011) it
occurs at a time ranging from 350~My to over 1~Gy for disks with inner
edge ranging from 15.5 to 20~AU (Neptune is at $\sim 11.5$~AU in these
simulations), with no apparent correlation between instability time
and initial location of the inner edge of the disk.

Together, the papers by Morbidelli et al. (2007) and Levison et
al. (2011) build the new version of the ``Nice model''. This is much
superior than its original version (Tsiganis et al., 2005; Gomes et
al., 2005) because
(i) it removes the arbitrary character of the initial conditions of
the planets by adopting as initial configuration one of the end-states
of hydro-dynamical simulations and (ii) it removes the sensitive
dependence of the instability time on the location of the inner edge
of the disk; instead, a late instability seems to be a generic
outcome.

\section{Conclusions}

According to our new understanding  the evolution of the Solar System was
characterized by three main ``eras'': In the {\it gas-disk era}, the giant
planets acquired a multi-resonant configuration, in which each planet
was in a mean-motion resonance with its neighbor. Given the
Jupiter/Saturn mass-ratio, this prevented further inward migration,
and explains why Jupiter was not closer than 5 AU from the Sun at the
disappearance of the gas.  It is possible that the giant planets had
an inward-then-outward migration, bringing Jupiter temporarily at
$\sim 1.5$ AU (Walsh et al., 2010).  At the disappearance of the gas,
the system entered in the {\it planetesimal-disk era}. The orbits of the
giant planets were at the time much closer to each other than they are
now, and had significantly smaller eccentricities and
inclinations.  A massive disk of
planetesimals persisted outside the orbit of the outermost giant
planet.  The gravitational interactions between the giant
planets and this disk, slowly modified the resonant orbit of the
former. Eventually, $\sim 600$~My later, the giant planets became
unstable, as a result of these slow orbital modifications. The chaotic
phase that followed reshuffled the structure of the outer Solar
System: the giant planets acquired their current orbits; most of the
distant planetesimal disk was dispersed, causing the Late Heavy
Bombardment of the terrestrial planets; a small fraction of the
distant planetesimals got stranded in what we call today the Kuiper
belt.  With this profound re-organization, the Solar System entered
into the {\it current era}, lasting since $\sim 3.8$~Gy ago, in which it did
not suffer any further significant change.

This is a radically different view with respect to the one that was
consensual even just 10 years ago. However, it has a level of internal
coherence and a consistency with the observed structure of the Solar
System that have never been achieved before. This model describes a
Solar System evolving under the same two main processes usually
invoked to explain the structure of extra-solar planetary systems:
radial migration in the gas-disk and global orbital instability. In fact,
the simulations of the new Nice model, when they fail to reproduce our own
system, often lead to
planetary systems similar to some of those observed around other
stars, with very eccentric planets or planets that remain in resonance
forever. Thus, the great diversity among planetary systems seems to
stem not from a diversity of processes, but from the diversity of
outcomes under the same processes. This is due to the extreme sensitivity of
the evolution to the initial and environmental conditions.

Nevertheless, the Nice story is not complete yet. It needs to be
complemented with a model of the accretion of the giant planets that
is consistent with their inferred dynamical evolution, which still
does not exist.

\section{References}

\begin{itemize}

\item[$\bullet$] F. C. Adams et al. Photoevaporation of Circumstellar Disks 
Due to External Far-Ultraviolet Radiation in Stellar Aggregates.\ The 
Astrophysical Journal 611 (2004) 360-379. 
\item[$\bullet$]  J.C. Augereau, 
H. Beust.\ On the AU Microscopii debris disk. Density profiles, 
grain properties, and dust dynamics.\ Astronomy and Astrophysics 455 (2006)
987-999. 
\item[$\bullet$] K. Batygin, M.E. Brown, 
 Early Dynamical Evolution of the Solar System: Pinning Down 
the Initial Condition of the Nice Model.The Astrophysical Journal 716 (2010)
1323-1331. 
\item[$\bullet$] W.F. Bottke et al. The 
Irregular Satellites: The Most Collisionally Evolved Populations in the 
Solar System.\ The Astronomical Journal 139 (2010) 994-1014. 
\item[$\bullet$]  R. Brasser et al.  Constructing the secular
  architecture of the Solar System II: the terrestrial
  planets. Astronomy and Astrophysics 507 (2009) 1053-1065.
\item[$\bullet$] E.I. Chiang. 
Excitation of Orbital Eccentricities by Repeated Resonance Crossings: 
Requirements.\ The Astrophysical Journal 584 (2003) 465-471.
\item[$\bullet$] E.I. Chiang, Y.  
Lithwick. Neptune Trojans as a Test Bed for Planet Formation.\ 
The Astrophysical Journal 628 (2005) 520-532. 
\item[$\bullet$] M. {\'C}uk, J.A. Burns. 
 Gas-drag-assisted capture of Himalia's family.\ Icarus 167 (2004) 
369-381. 
\item[$\bullet$] L. Dones et al. in  M. Festou, H. A. Weaver, 
  F. Keller (Eds.), {Comets II},
University Arizona Press., Tucson, 2004, p. 153-174. 
\item[$\bullet$] M.J. Duncan, H.F. Levison. Scattered comet disk and the
  origin of Jupiter family comets. {Science} 276 (1997) 1670--1672.
\item[$\bullet$] J.A. Fernandez, W.H. Ip. Some dynamical aspects of the
  accretion of Uranus and Neptune - The exchange of orbital angular
  momentum with planetesimals. {Icarus} 58 (1984), 109-120.
\item[$\bullet$] R.S. Gomes. Dynamical 
Effects of Planetary Migration on Primordial Trojan-Type Asteroids.\ The 
Astronomical Journal 116 (1998) 2590-2597. 
\item[$\bullet$] R.S. Gomes, A. Morbidelli, 
H.F. Levison. Planetary migration in a planetesimal disk: why 
did Neptune stop at 30 AU?.\ Icarus 170 (2004), 492-507. 
\item[$\bullet$] R. Gomes et al.  Origin of the cataclysmic Late Heavy
Bombardment period of the terrestrial planets. Nature 435 (2005)
466-469.
\item[$\bullet$] J.M. Hahn, R. Malhotra. Orbital Evolution of Planets
  Embedded in a Planetesimal Disk. {Astronomical Journal} 117 (1999),
  3041-3053.
\item[$\bullet$] T.A. Heppenheimer, 
C. Porco. New contributions to the problem of capture.\ 
Icarus 30 (1977) 385-401. 
\item[$\bullet$] D.C. Jewitt, 
S.S. Sheppard. Physical Properties of Trans-Neptunian Object 
(20000) Varuna.\ Astronomical Journal 123 (2002) 2110-2120.
\item[$\bullet$] S.J. Kenyon, B.C. 
Bromley.\ Stellar encounters as the origin of distant Solar 
System objects in highly eccentric orbits.\ Nature 432 (2004) 598-602. 
\item[$\bullet$] S.J. Kortenkamp.
An efficient, low-velocity, resonant mechanism for capture of satellites by 
a protoplanet.\ Icarus 175 (2005) 409-418. 
\item[$\bullet$] J.L. Lagrange. Essai sur le problème des trois
  corps. Oeuvres de Lagrange, Tome 6, Chapitre
  II. (1787) Gauthier-Villars. pp. 272-292.
\item[$\bullet$] P.S. Laplace. Expositions du syst\`eme du
  monde. (1796) Imprimerie Cercle-Social, Paris. 
\item[$\bullet$] H.F. Levison et al.  Could 
the Lunar ``Late Heavy Bombardment'' Have Been Triggered by the Formation 
of Uranus and Neptune?.\ Icarus 151 (2001) 286-306. 
\item[$\bullet$] H.F. Levison et al. in  Protostars and Planets V,
  University Arizona Press, Tucson, 2007, p. 669-684.
\item[$\bullet$]  H.F.Levison et al. Origin of 
the structure of the Kuiper belt during a dynamical instability in the 
orbits of Uranus and Neptune.\ Icarus 196 (2008) 258-273. 
\item[$\bullet$] H.F. Levison et al. Reevaluating the Early Dynamical Evolution of the Outer Planets.  in preparation.
\item[$\bullet$] F. Marzari et al. in W.F. Bottke et al. (Eds.)
Asteroids III, University of Arizona Press., Tucson, 2002, p. 725-738.
\item[$\bullet$] F. Masset, M.
Snellgrove. Reversing type II migration: resonance trapping of a 
lighter giant protoplanet. Monthly Notices of the Royal Astronomical 
Society 320 (2001) 55-59. 
\item[$\bullet$] T.A. Michtchenko, C. Beaug{\' e}, F. Roig. Planetary
  Migration and the Effects of Mean Motion Resonances on Jupiter's Trojan
  Asteroids.\ Astronomical Journal 122 (2001) 3485-3491 
\item[$\bullet$] A. Morbidelli et al. Chaotic capture of 
Jupiter's Trojan asteroids in the early Solar System.\ Nature 435 (2005)
462-465.
\item[$\bullet$] A. Morbidelli, A. 
Crida. The dynamics of Jupiter and Saturn in the gaseous 
protoplanetary disk. Icarus 191 (2007) 158-171. 
\item[$\bullet$] A. Morbidelli et al.  Dynamics of the 
Giant Planets of the Solar System in the Gaseous Protoplanetary Disk and 
Their Relationship to the Current Orbital Architecture.\ The Astronomical 
Journal 134 (2007) 1790-1798. 
\item[$\bullet$] A. Morbidelli et al. Constructing the secular
  architecture of the Solar System. I. The giant planets.\ Astronomy
  and Astrophysics 507 (2009) 1041-1052. 
\item[$\bullet$] A. Morbidelli et al.  Evidence from the asteroid belt for a
  violent past evolution of Jupiter's orbit. Astron. J. (2010) in press. 
\item[$\bullet$] D. Nesvorn{\'y}, D. 
Vokrouhlick{\'y}, A. Morbidelli.\ Capture of Irregular 
Satellites during Planetary Encounters.\ The Astronomical Journal 133 (2007) 
1962-1976. 
\item[$\bullet$] D.P. O'Brien, A. 
Morbidelli, W.F. Bottke. The primordial excitation and 
clearing of the asteroid belt - Revisited.\ Icarus 191 (2007) 434-452.
\item[$\bullet$]  A. Pierens, R.P. Nelson. Constraints on
  resonant-trapping for two planets embedded in a protoplanetary
  disc.\ Astronomy and Astrophysics 482 (2008) 333-340.
\item[$\bullet$] E.W. Thommes, M.J. Duncan, H.F. Levison. The formation of Uranus and Neptune in the Jupiter-Saturn region of the Solar 
System. Nature 402 (1999) 635-638.
\item[$\bullet$] K. Tsiganis et al. Origin of the orbital 
architecture of the giant planets of the Solar System. Nature 435 (2005) 
459-461. 
\item[$\bullet$] A.N. Youdin, J. 
Goodman. Streaming Instabilities in Protoplanetary Disks.\ The 
Astrophysical Journal 620 (2005) 459-469. 
\item[$\bullet$] K. Walsh et al. Origin of the Asteroid Belt and Mars'
  Small Mass. DPS (2010) 04.02
\item[$\bullet$]  G.W. Wetherill,
G.R., Stewart. Formation of planetary embryos - Effects of
fragmentation, low relative velocity, and independent variation of
eccentricity and inclination.\ {\it Icarus} 106 (1993), 190-201.
\item[$\bullet$] S.J. Weidenschilling. The distribution of mass in the
  planetary system and solar nebula.\ Astrophysics and Space Science
  51 (1977) 153-158.

\end{itemize}

\end{document}